\documentclass{sf2a-conf2014}
\usepackage{graphicx}
\usepackage{hyperref}
\usepackage[]{natbib}  
\usepackage{epstopdf}

\def\BibTeX{{\rm B\kern-.05em{\sc i\kern-.025em b}\kern-.08em
    T\kern-.1667em\lower.7ex\hbox{E}\kern-.125emX}}
\bibpunct{(}{)}{;}{a}{}{,}  

\begin{document}

\TitreGlobal{SF2A 2014}


\title{Helioseismic inferences of the solar cycles 23 and 24:\\
GOLF and VIRGO observations}

\runningtitle{Helioseismic inferences of the solar cycles 23 and 24}

\author{D. Salabert}\address{Laboratoire AIM, CEA/DSM-CNRS, Universit\'e Paris 7 Diderot, IRFU/SAp, Centre de Saclay, 91191 Gif-sur-Yvette, France}
\author{R.A. Garc\'ia$^1$}
\author{A. Jim\'enez$^{2,}$}\address{Instituto de Astrof\'isica de Canarias, 38200 La Laguna, Tenerife, Spain}\address{Departemento de Astrof\'isica, Universidad de La Laguna, 38205 La Laguna, Tenerife, Spain}

\setcounter{page}{237}


\maketitle


\begin{abstract}
The Sun-as-a star helioseismic spectrophotometer GOLF and photometer VIRGO instruments onboard the SoHO spacecraft are collecting high-quality, continuous data since April 1996. We analyze here these unique datasets in order to investigate the peculiar and weak on-going solar cycle 24. As this cycle 24 is reaching its maximum, we compare its rising phase with the rising phase of the previous solar cycle 23.
\end{abstract}



\section{Introduction}
 As now very well established, the acoustic frequencies vary along the solar cycle and show high levels of correlation with the solar activity proxies. However, in this work, we find evidences of significant differences in the frequency dependence of the acoustic frequency variations between the two rising phases of cycles 23 and 24, providing insights into the related changes in magnetic field and internal structure.
 
\section{Data and analysis}
We analyzed simultaneous space-based, Sun-as-a-star helioseismic observations collected by the Global Oscillations at Low Frequency \citep[GOLF;][]{gabriel95} and the Variability of Solar Irradiance and Gravity Oscillations  \citep[VIRGO;][]{froh95} instruments onboard the {\it Solar and Heliospheric Observatory} (SoHO) spacecraft.
GOLF measures the radial velocity Doppler shift - integrated over the solar surface - in the D1 and D2 Fraunhofer sodium lines at 589.6 and 589.0 nm respectively. VIRGO is composed of three Sun photometers (SPM) at 402 nm (Blue), 500 nm (Green) and 862 nm (Red). The GOLF velocity time series were obtained following \citet{garcia05} and calibrated as described in \citet{chano03}, while the VIRGO photometric observations were calibrated as described in \citet{jimenez02}.
These two instruments are providing unique datasets of high-quality, continuous observations of the low-degree oscillation modes since April 1996 (i.e. since more than 18 years today) covering the solar activity cycles 23 and 24. A total of 6538 days of observations were analyzed, spanning the period from 1996 April 11 to 2014 March 5, with an overall duty cycle larger than 96\%.\\

These two datasets were split into contiguous 730-day, 365-day, and 182.5-day sub series, each series being allowed to overlap by 182.5 days, 91.25 days, and 46.625 days respectively. The power spectrum of each sub series was fitted to estimate the $l = 0$, 1, 2, and 3 mode parameters using a standard likelihood maximization function as described in \citet{salabert07}.  Each mode component was parameterized using an asymmetric Lorentzian profile \citep{nigam98}. The amplitude ratios between the $l=0,1,2$, and 3 modes and the $m$-height ratios of the $l=2$ and 3 multiplets calculated in \citet{salabert11} for the GOLF and VIRGO measurements were used.  
The $l = 4$ and  $l = 5$ were also included in the fitted profile when present in the fitted window. 

\section{Temporal variations of the low-degree oscillation frequencies over 6500 days}
\begin{figure}[ht!]
 \centering
 \includegraphics[width=1\textwidth]{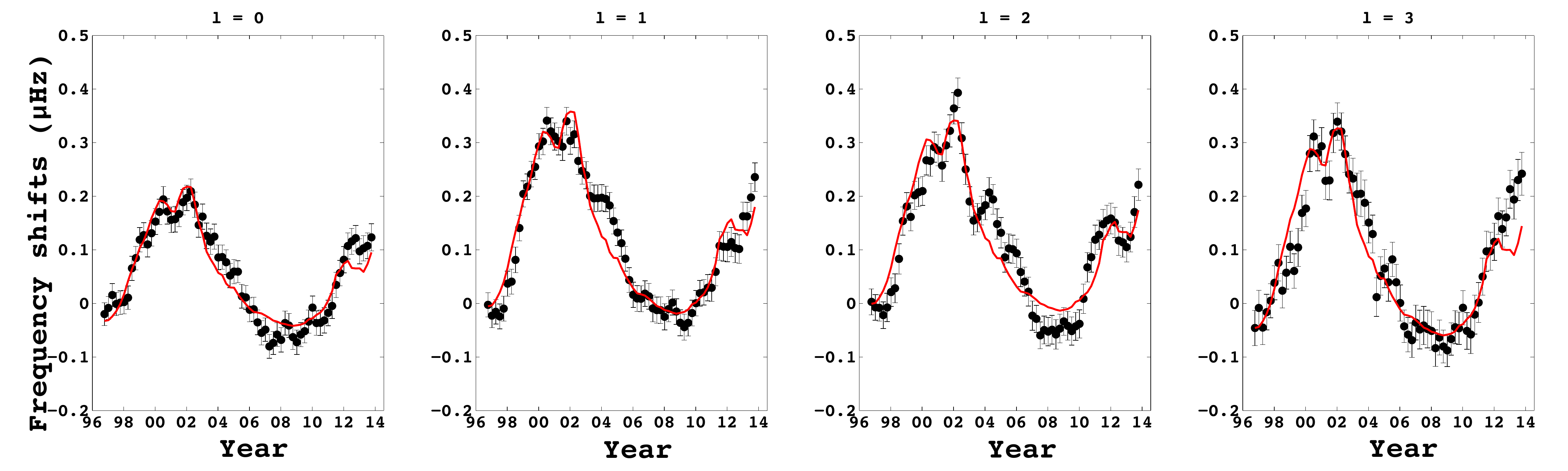}      
  \caption{Temporal variations of the frequency shifts in $\mu$Hz of the  individual $l=0$, 1, 2, and 3 modes (from left to right respectively) and averaged between 2450~$\mu$Hz and 3520~$\mu$Hz extracted from the analysis of the 365-day GOLF spectra (black dots). The scaled 10.7-cm radio flux, $F_{10.7}$, averaged over the same 365-day timespan is shown as a proxy of the solar surface activity (solid lines).}
  \label{fig:fig1}
\end{figure}

The temporal variations of the p-mode frequencies were defined as the differences between the mode frequencies observed at different dates and reference values of the corresponding modes taken as the average over the years 1996-1997 during the minimum of cycle 22. The formal uncertainties returned from the peak-fitting analysis were used as weights in the average computation. In addition, mean values of daily measurements of the 10.7-cm radio flux, $F_{10.7}$, were obtained and used as a proxy of the solar surface activity. The frequency shifts measured at each individual angular degree, $l = 0$, 1, 2, and 3, and averaged between 2450~$\mu$Hz and 3520~$\mu$Hz extracted from the GOLF 365-day sub series are shown in Fig.~\ref{fig:fig1}. The corresponding scaled 10.7-cm radio flux is also represented by the red solid lines. The signature of the quasi-biennal oscillation \citep{fletcher10} is also clearly observable in Fig.~\ref{fig:fig1}. The GOLF and VIRGO frequency shifts per unit of change in the 10.7-cm radio flux in nHz/RF\footnote{The 10.7-cm radio flux has for units 1~RF =  10$^{-22}$~J~s$^{-1}$~m$^{-2}$~Hz$^{-1}$.} calculated through weighed linear regressions between these two quantities, and the associated linear correlation coefficients are given in Table~\ref{table:correl}.

\begin{table*}[ht]
\begin{minipage}{\textwidth}
\caption{Variations of the frequencies shifts at each $l$ per unit of change in the 10.7-cm radio flux (nHz/RF) and associated linear correlations, $r_p$,  obtained from the analysis of the 365-day GOLF and VIRGO spectra. The frequency shifts were calculated between 2450~$\mu$Hz and 3520~$\mu$Hz.  Independent points only were used. }
\label{table:correl}      
\centering               
\renewcommand{\footnoterule}{}  
\begin{tabular}{c c c c c }        
\hline\hline   
& \multicolumn{2}{c}{GOLF} & \multicolumn{2}{c}{VIRGO}\\ 
 $l$  &Gradient\footnote{Gradient against the 10.7-cm radio flux in units of nHz RF$^{-1}$ (with 1~RF =  10$^{-22}$~J~s$^{-1}$~m$^{-2}$~Hz$^{-1}$).} & $r_p$  &Gradient$^a$ & $r_p$ \\    
\hline                           
 
0  & 2.07~$\pm$~0.14 & 0.95 & 2.55~$\pm$~0.10 & 0.95  \\
1  & 3.04~$\pm$~0.15 & 0.97 & 3.54~$\pm$~0.11 & 0.96  \\
2  & 2.85~$\pm$~0.17 & 0.94 & 2.92~$\pm$~0.13 & 0.95  \\
2  & 3.10~$\pm$~0.21 & 0.94 & - & -  \\

 \hline                                
\end{tabular}
\end{minipage}
\end{table*}

\begin{figure}[ht!]
 \centering
 \includegraphics[width=0.99\textwidth,clip]{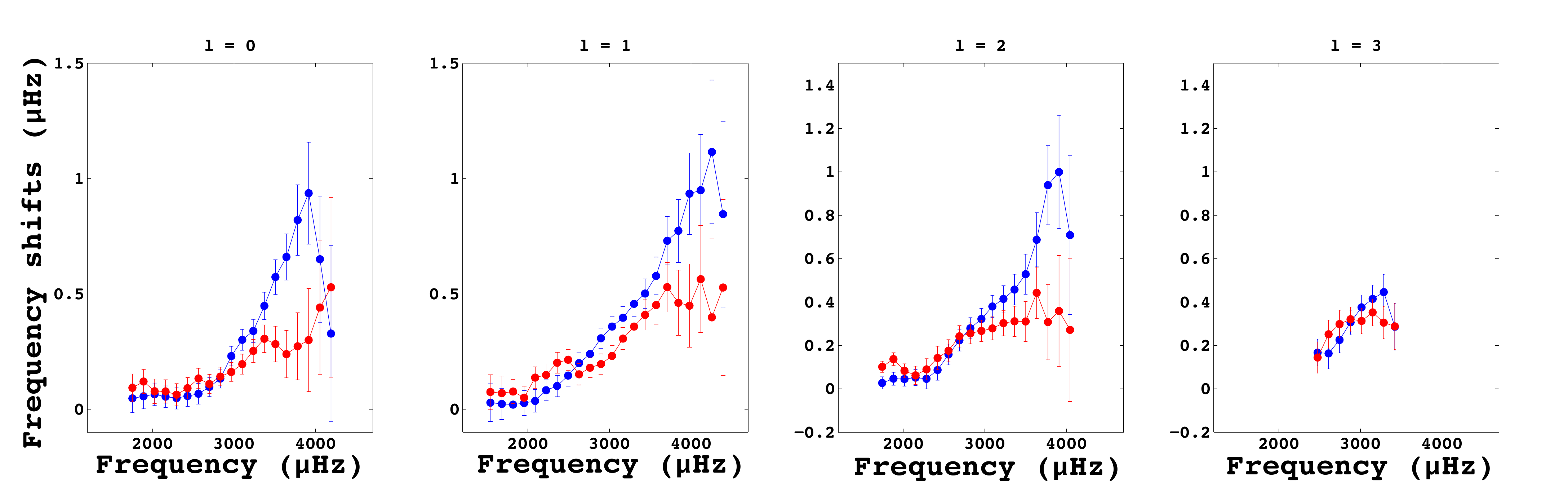}\\     
 \includegraphics[width=0.99\textwidth,clip]{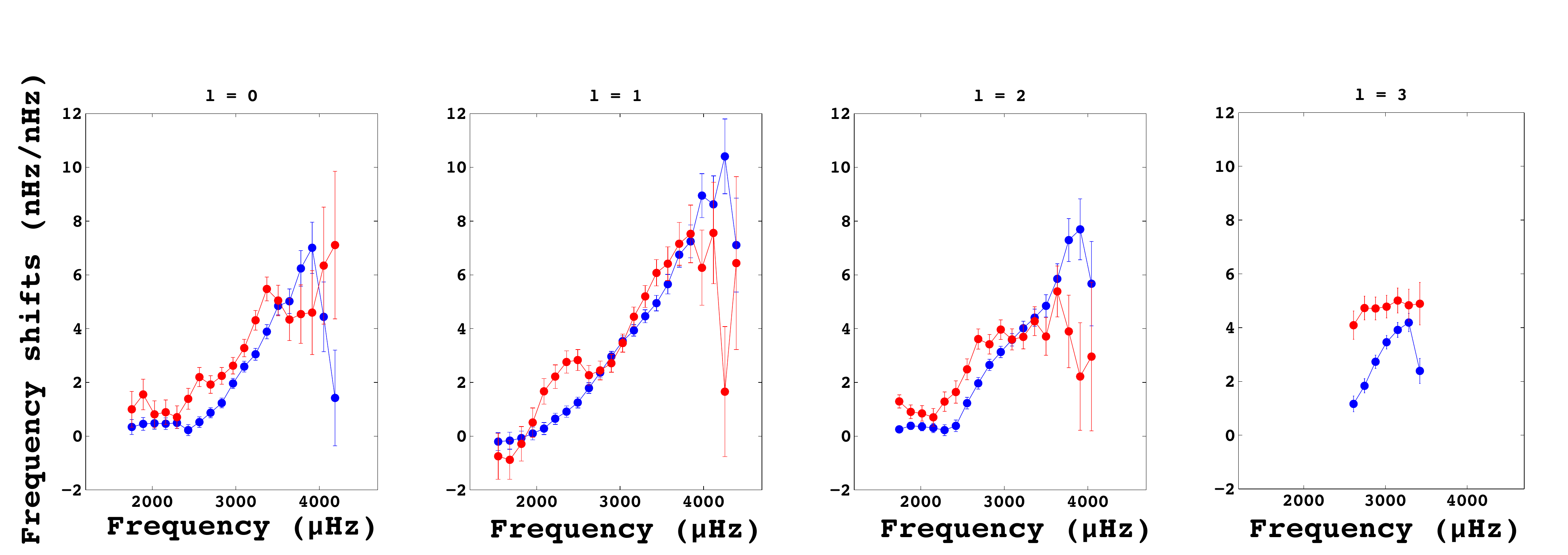}      
  \caption{{\bf Top:} Frequency shifts in $\mu$Hz of the individual $l = 0$, 1, 2, and 3 modes  (from left to right respectively) as a function of frequency during the rising phase of the solar cycle 23 (blue dots) and the rising phase of the solar cycle 24 (red dots). {\bf Bottom:} Same as the top panels but for the frequency shifts per unit of change in the 10.7-cm radio flux (nHz/RF).}
  \label{fig:fig2}
\end{figure}

In the following, we studied the frequency shifts as a function of frequency during the rising phases of cycles 23 and 24. 
The top panels of Fig.~\ref{fig:fig2} show the individual $l=0$, 1, 2, and 3  frequency shifts as a function of frequency during the rising phases of cycles 23 (blue) and 24 (red) extracted from the GOLF observations. The 2-year periodicity was removed by applying a proper smoothing as explained in \citet{fletcher10}.
Differences in the frequency shifts are observed between the two cycles 23 and 24. First, the low-frequency part of the p-mode oscillation power spectrum, which is more sensitive to deeper sub-surface layers of the Sun, show similar shifts between the two cycles. On the other hand, in the high-frequency range of the power spectrum, where the shifts are larger, the variations of the frequency shifts are smaller during the rising phase of the solar cycle 24. As the frequency shifts at higher frequencies are highly correlated with surface activity, these smaller shifts at high frequency are consistent with the weaker surface activity observed during cycle 24.
The bottom panels of Fig.~\ref{fig:fig2} show the frequency shifts per unit of change in the 10.7-cm radio flux as a function of frequency during the rising phases of cycles 23 (blue) and 24 (red) extracted from the GOLF observations. The frequency variations per unit of surface activity are larger in cycle 24 than in cycle 23. The results obtained from the VIRGO data are consistent within the error bars with GOLF. \\

\begin{figure}[ht!]
 \centering
 \includegraphics[width=0.5\textwidth]{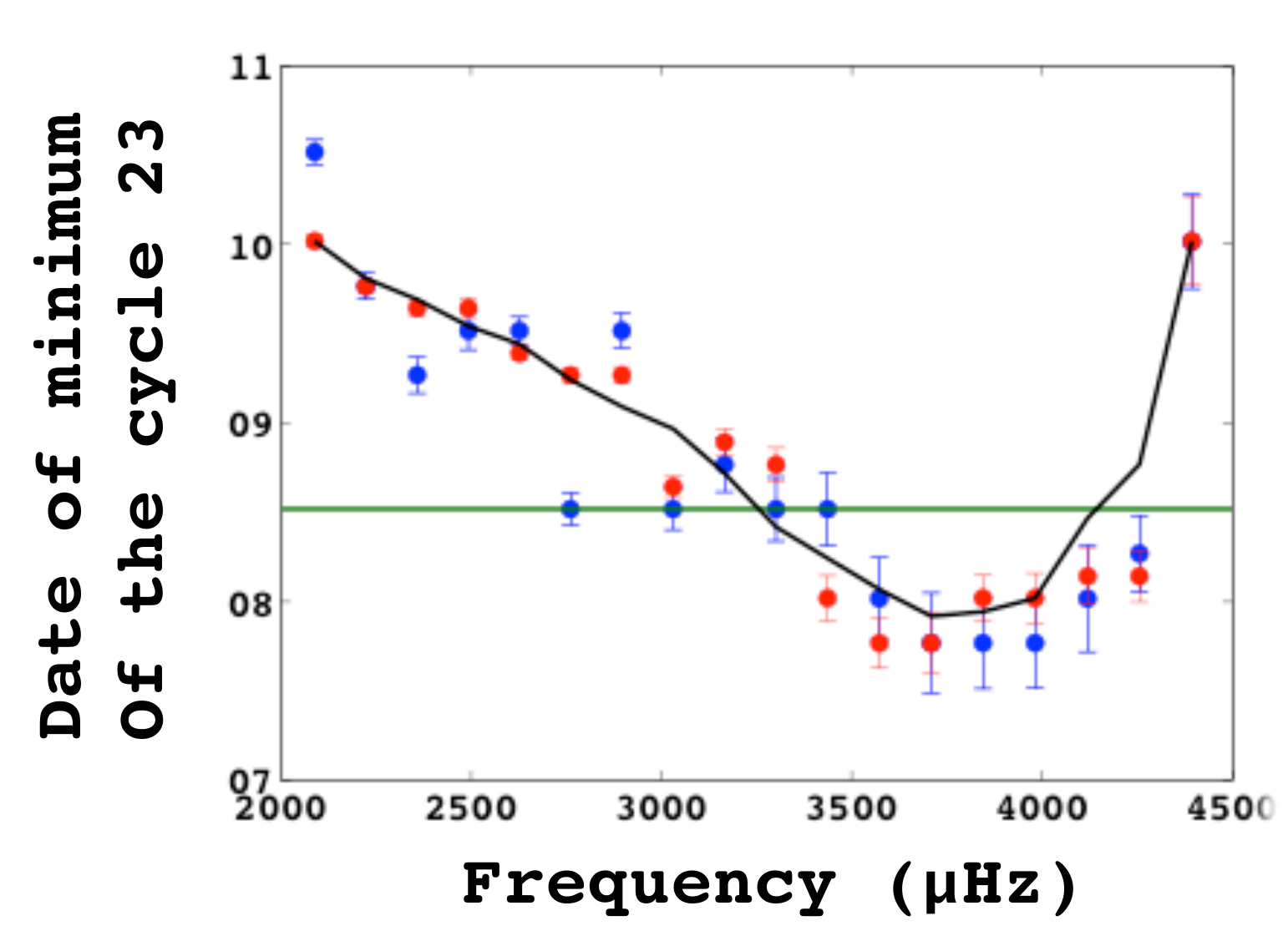}      
  \caption{Estimated dates of the minimum of the temporal variations of the $l=1$ frequency shifts associated to solar activity during the minimum of the solar cycle 23 (2008-2009). In blue, the dates obtained from the 365-day sub series, and in red, the ones obtained from the 182.5-day sub series. The solid black line corresponds to the smoothed 182.5-day results. The horizontal green solid line corresponds to the minimum of the smoothed 10.7-cm radio flux.}
  \label{fig:fig3}
\end{figure}

Figure~\ref{fig:fig3} shows the estimated dates of the minimum of cycle 23 (2008-2009) as estimated by the $l=1$ mode frequency shifts by measuring the minimum of their temporal variations  as a function of frequency. These dates were calculated after removing the signature of the 2-year periodicity. The results extracted from the analysis of the 365-day GOLF spectra are represented in blue, while the results extracted from the analysis of the 182.5-day GOLF spectra are represented in red. The solid black line corresponds to the smoothed 182.5-day results. The horizontal green solid line corresponds to the minimum of cycle 23 measured from the smoothed 10.7-cm radio flux. The minima of the temporal variations of the mode frequency are different from the minimum estimated based on the 10.7-cm radio flux.

\section{Conclusions}
We analyzed more than 6500 days of radial velocity GOLF and photometric VIRGO observations. We found differences in the temporal variations of the low-degree oscillation frequencies between the two rising phases of cycles 23 and 24 with an important frequency dependence. The results obtained in this study suggest that the solar magnetic field and internal structure in the upper sub-surface layers of the Sun have changed between cycle 23 and cycle 24. It would indicate as well that in deeper layers  inside the Sun, magnetic field and internal structure are similar between cycle 23 and cycle 24. More precise localization of where the changes are taking place will need the analysis of high-degree modes.

\begin{acknowledgements}
The GOLF and VIRGO instruments onboard SoHO are a cooperative effort of many individuals, to whom we are indebted. SoHO is a project of international collaboration between ESA and NASA. D.S. and R.A.G. acknowledge the support from the CNES/GOLF grant at the SAp CEA-Saclay. The 10.7-cm solar radio flux data were obtained from the National Geophysical Data Center.
\end{acknowledgements}

\end{document}